\documentclass[useAMS,usenatbib]{nature}

\usepackage{amsmath}
\usepackage{pdflscape}
\usepackage{afterpage}

\usepackage{epsfig}
\usepackage{epstopdf}
\usepackage{lscape} 
\usepackage{natbib}
\usepackage{tabularx}
\usepackage{multirow}
\usepackage{amssymb}
\def\gtrsim{\mathrel{\hbox{\rlap{\hbox{\lower4pt\hbox{$\sim$}}}\hbox{$>$}}}}
\newcommand{\apj}{Astrophys. J.}       
\newcommand{\apjl}{Astrophys. J. Lett.}      
\newcommand{\mnras}{Mon. Not. R. Astron. Soc.}   
\newcommand{\nat}{Nature}       
\newcommand{\pasj}{Publ. Astron. Soc. Japan}     
\newcommand{\aap}{Astron. Astrophys.}
\newcommand{\prd}{Phys. Rev. D}

\title{What are Fast Radio Bursts?}
\author{
Ue-Li Pen $^{1*}$
\\
$^1$ Canadian Institute for Theoretical Astrophysics, University of Toronto, Ontario, Canada. \\
$*$E-mail:\ pen@cita.utoronto.ca
}
\begin{document}
\date{\today}
\maketitle
\label{firstpage}

\begin{abstract}
Physical constraints on the sources of Fast Radio Bursts are few, and therefore viable theoretical models are many. However, no one model can match all the available observational characteristics, meaning that these radio bursts remain one of astrophysics' most mysterious phenomena.
\end{abstract}

\newcommand{\be}{\begin{eqnarray}}
\newcommand{\ee}{\end{eqnarray}}
\newcommand{\beq}{\begin{equation}}
\newcommand{\eeq}{\end{equation}}

\section{Introduction}

Fast Radio Bursts (FRBs) are a relatively new astrophysical mystery.  Their novelty means that the field is rapidly changing, and some aspects of this article may become outdated in the month from submission to publication.

A large number of models for FRB progenitors have been proposed, many more than can be
readily summarized in this Comment.  I present the current collection of
physical constraints, as well as some basic caveats, and compare a
representative sample of models against these.  Many models only aim
to explain a subset of evidence (for example, cataclysmic models will
not repeat, and therefore will fail to explain FRB 121102, the only known repeating FRB), but given the early days of the field, I allow
consideration of the possibility that more than one mechanism is at
play.  I summarize with a forward look, pointing to upcoming promising
lines of evidence.

FRBs are defined as radio-frequency impulses more highly dispersed than allowed by the Milky
Way's diffuse ionized medium \citep{2007Sci...318..777L,
  2013Sci...341...53T}, and this dispersion is quantified in the dispersion measure (DM), measured in parsecs per cubic centimetre.  The excess DM (over that imposed by the Milky Way) is presumably due to the
immediate vicinity of the FRB, or propagation through the
intergalactic medium (IGM).  The IGM contribution cannot be larger than the
observed DM, placing an upper bound on their distance $D$, and thus a
lower bound on their spatial number density.  Their radial position
could be anything from terrestrial to cosmological
\citep{2014ApJ...797...70K}.  The inferred space density is high
if a large fraction of the DM is physically associated with FRBs,
which I call the local scenario: $D \sim$ 10-2,000 Mpc.  In these
scenarios, the DM could be dominated by dense plasmas, including
supernova remnants, H\,{\sc ii} regions, circumnuclear gas, or stellar
outflows/atmospheres.  Placing FRBs much closer would result in their
event rate being dominated by the nearest neighbouring galaxies, while more
distant distributions are likely affected by cosmological evolution.
The nearby scenarios share Euclidean statistics\citep{1968ApJ...151..393S}: as one increases the sensitivity
by a factor of 4, one probes to twice the distance, or eight times the
volume, and expects eight times the event rate (analogous to Olber's
paradox), and sources are described by a flux count slope $d\ln N/d\ln S$ of -1.5, where $N$ denotes the number of bursts of flux $S$. 
This slope should be uncorrelated with any other properties, e.g. DM.  If a correlation between DM and flux is found, it would rule out all local
models.

The majority of information (and bursts) come from a single object,
FRB121102, for which the distance, host galaxy, and repetition
statistics are known.  Its repetition is
non-Poissonian\citep{2018MNRAS.475.5109O}, and it is certainly
possible that all FRBs function by the same mechanism (see the Comment by Manisha Caleb et al. in this issue).  While some FRBs have been followed up extensively without detection of repetition, there
also appear to be extended periods of months where FRB121102 does not
repeat.

Extrapolating from a single event to a population results in large
uncertainties. FRB121102 is consistent with both local and
cosmological scenarios.  Being the only event detected by Arecibo Observatory, which searched a small area of sky very deeply, in a local scenario it could easily be the furthest of all bursts.

\section{Constraints on FRB sources}

{\bf Energetics.} the total energy of FRBs is modest: the typical
energy of 1 Jy-ms at 1 GHz for a distance of 1 Gpc is $10^{30}$~J, comparable to the Crab Pulsar spin-down energy, and smaller than soft gamma-ray outbursts.  All FRB models are built to explain the basic energetics.  The harder challenge is the emission in gigahertz radio waves: most mechanisms propose an energy injection with an unknown conversion process, with the notable exception of primordial dipoles.

{\bf Coherence.} A separate challenging aspect of FRB mechanisms is
the high observed brightness temperature.  If the physical size of the source is no
larger than the apparent duration of the event, this corresponds to a
range of a few to thousands of kilometres.  Converting the radiation
field to an equivalent brightness temperature yields $10^{40+}$ K,
higher than the Planck Temperature!

The same challenge exists for short duration giant pulses, or
nanoshots, oberversed in pulsars.  Presumably the emission region is
not thermal, but rather consists of coherent motions of electrons: the
brightness temperature of an FM radio station is $10^{32}$K, also
comparable to the Planck temperature. This is a statement that all the
electrons are moving coherently, and not in a thermally random fashion.
This precludes synchrotron emission.
The coherent conversion poses a physical challenge to any mechanism, somewhat analogous to the gamma-ray burst non-thermal constraint.  Very few models propose a quantitative physical mechanism to achieve this.

{\bf Additional constraints.} FRBs all appear to scintillate, similar to
pulsars.  Almost no other radio sources are compact enough ($\lesssim$
nano-arcseconds) enough to scintillate on megahertz frequency scales from
plasma lenses in the Milky Way.  FRBs exhibit frequency structures
that are not power laws.  This could be due to plasma lensing near the
source\citep{2017ApJ...842...35C,2018Natur.557..522M}, or intrinsic to
the source.  The latter is challenging to understand: the energy
densities involved require relativistic degrees of freedom, which
tends to result in a spread of frequencies due to the relativistic
Doppler shifts. The superradiance proposal (see below) is an
exception, which generically predicts frequency structure. Any
characteristic resonant length scale would need to be small, on the order of a metre, which does not have any known physical candidate mechanism.
An analogous mystery is the gigahertz banding structure in Crab interpulses
(for recent summary, see \cite{2016JPlPh..82c6302E}).

{\bf Magnetic properties:} Some FRBs are highly polarized, led by FRBs 110523 and 150215\citep{2017MNRAS.469.4465P}.  For many events, polarization information was not recorded, and the high RM ($\sim 10^5$) observed in the Repeater would have been bandwidth smeared, so it is possible that all FRBs are highly polarized.  The RM varies from small to very large, and the larger values are imprinted near the source.  The only other line of sight known to lead to a comparably high RM is the Galactic Centre, with magnetar J1745-2900 showing an RM$\sim 10^5$ as well\citep{2013Natur.501..391E}.  Both it and FRB121102 show non-monotonic RM variations on timescales of a month.

The above properties are ideally explained by model(s) of FRBs.  Most
models only attempt to address a subset of the known constraints, but many
are at odds with the collective data.  This would be expected if more than one mechanism of generating FRBs exists.

\section{Models of FRB progenitors}

I list a selection of models in decreasing order of mundaneness:

{\bf Radio frequency interference (RFI):} terrestrial signals had initially been a concern, and some events (perytons) were indeed traced back to a microwave oven\citep{2015MNRAS.451.3933P}.  VLBI detection ruled out interference for the Repeater, as it is unthinkable for the voltages to correlate across continents.  Perytons serve as a concrete example of an identified signal that shares similarities with FRBs, demonstrating the
possibility of multiple populations.

{\bf Flare Stars:} some stars are known to exhibit short time variation
radio emission in radio flares, making them a potential FRB
candidate\citep{2014MNRAS.439L..46L}.  These stars might be in the Galactic halo, with the DM imprinted in the stellar corona.  Models
predict some deviation from the $\lambda^2$ dispersion law, and cannot
explain the few bursts where the wavelength dependence has been precisely measured.

{\bf Neutron star/pulsar subcategories}:
Young pulsar giant pulses: Maxim Lyutikov discussed the
energetics\citep{2016MNRAS.462..941L}, which can be met in a local
scenario. Magnetars have substantial energy stored in magnetic fields,
which might be released.  The repeating burst shows a number of
similarities with the Galactic Centre
magnetar\citep{2015ApJ...807..179P,2018arXiv180809969M}.  Objects may
hit pulsars\citep{2014ApJ...780L..31B}, they can
quake\citep{2018ApJ...852..140W} and transition between phases
\citep{2016RAA....16...80S}.  The compact nature of neutron
stars and the \textbf{known unknown} mechanism of their radio emission makes them popular, though mostly non-predictive.

Blitzars:  the collapse of a neutron star into a black hole is called
a Blitzar\citep{2014A&A...562A.137F}. Magnetic fields have to release their energy since black holes can't have hair, and may accelerate electrons to produce the radio bursts.

{\bf Black hole accretion flows:} it is not surprising that the energetics
near black holes has been invoked to power FRBs\citep{2017A&A...602A..64V}. The short duration can be associated with a wandering beam, analogous to shining a laser pointer into space.

{\bf Superradiance:} a version of maser emission, this postulates a radio
line in a relativistic disk orbiting a black hole to smear into the broad
observed radio features\citep{2018MNRAS.475..514H}.  This and the
following model are one of the very few which describe a quantifiable
coherent emission mechanism.

{\bf Dark matter:} some models of magnetic dipole dark
matter\citep{2017ApJ...844..162T} predict many of the subsequently
observed properties, including irregular repetition, a flat spectral
index and a high RM.

{\bf Black hole explosions:} In the generic Hawking picture, primordial
black holes emit their final gasp when they reach the radius (or inverse
temperature) of massive particles, e.g. electrons.  One would expect
highly energetic gamma-rays to emit.  Some proposals increase this
radius using virtual white holes\citep{2014PhRvD..90l7503B} to explode
with $\sim$centimetre sizes, thus leading to FRBs.

{\bf Cosmic strings:} it has been proposed that cosmic string loop cusps can result in radio emission\citep{2017arXiv170702397B}.  While these do
not quantitatively explain the coherent spectrum of FRBs, they do predict
non-association with galaxies, which is rather unique.

{\bf Intelligent alien signals:} not sure how such proposals got through the refereeing process\citep{2014ApJ...785L..26L}, which surely have the
fewest robust predictions!

Combination of models can be, and have been, created using
combinations of these mechanisms, for example black holes and neutron
stars.  Many more models have been proposed, which this short Comment is unable to do justice to.  We summarize the models in Table 1.

\begin{landscape}
\begin{table*}
\scriptsize
\centering
\begin{tabularx}{1.08\textwidth}{@{\extracolsep{\fill}}|lccccccc|}
\hline
\multicolumn{1}{|c}{\textbf{Location}}
  & \textbf{Model}                                              &
                                                                  \textbf{\begin{tabular}[c]{@{}c@{}}
                                                                            repetition\end{tabular}}
  & \textbf{\begin{tabular}[c]{@{}c@{}}Faraday \\
              rotation\end{tabular}} &
                                       \textbf{$\mathbf{\frac{dlnN_{\rm FRB}}{dlnS_{\nu}}}$}                                      & \multicolumn{1}{l}{\textbf{Implications}}                                    & \textbf{\begin{tabular}[c]{@{}c@{}}DM range\\ (pc cm$^{-3}$)\end{tabular}} & \textbf{\begin{tabular}[c]{@{}c@{}}Scattering \end{tabular}} \\ \hline
\multicolumn{1}{|l|}{\multirow{4}{*}{\begin{tabular}[c]{@{}l@{}}Cosmological \\ ($\gtrsim 2 h^{-1}$Gpc)\end{tabular}}}             & blitzars          & $\times$            & $\lesssim 10$              & ?                                              & \begin{tabular}[c]{@{}c@{}}gravitational \\ waves\end{tabular}              & $10^{3-4}$                                                                & $\times$                                                                  \\
\multicolumn{1}{|l|}{}                                                                                                            & merging compact objects                                                 & $\times$            & $\lesssim 10$              & ?                                             & \begin{tabular}[c]{@{}c@{}}type Ia SNe,\\  X-ray, $\gamma$-ray\end{tabular} &           $10^{3-4}$                                                     & $\times$                                                                  \\

\multicolumn{1}{|l|}{}                                                                                                            & magnetar flare                                              & \checkmark                                                                       & $\lesssim 10$                                              & ?                                                                                      & \begin{tabular}[c]{@{}c@{}}$\sim$ms TeV \\ burst\end{tabular}               & $10^{3-4}$                                                                    & $\checkmark$                                                        \\ 
\multicolumn{1}{|l|}{}
  & dipole dark matter                                              &\checkmark          & $10^{3-5}$                                              & ?                                               & \begin{tabular}[c]{@{}c@{}}flat       radio spectrum\end{tabular}              &             $10^{3-4}$                                                      & $\checkmark$                                                        \\ 
\multicolumn{1}{|l|}{}
  & cosmic strings             &                                        $\times$  & $\lesssim 10$                                              & ?                                                    & \begin{tabular}[c]{@{}c@{}}no galaxy association     \end{tabular}        &   $10^{3-4}$                                                             & $\times$                                                        \\  \multicolumn{1}{|l|}{}
  & superradiance                                              &
                                                                  \checkmark
  & $10^{5-6}$                                              & ?
                                                                                                                                  &                                                                                                                    spectral regularity            & 300-2500                                                                & $\checkmark$                                                        \\ \cline{1-1}\multicolumn{1}{|l|}{\multirow{3}{*}{\begin{tabular}[c]{@{}l@{}}Extragalactic, local \\ ($\lesssim$2$ h^{-1}$Gpc)\end{tabular}}} & edge-on disk                                                & $\checkmark$                                                               & 50-500                                                      & -3/2                                                                                   &   \begin{tabular}[c]{@{}c@{}}scintillation \\resolvable      scattering\end{tabular}
  & 10-2000                                                                 & \checkmark                                                                 \\
\multicolumn{1}{|l|}{}
  & primordial BHs                                              &
                                                                  $\times$
  & $\lesssim 10$               & -3/2
                                                                                                                                  &  \begin{tabular}[c]{@{}c@{}}distributed\\
                                                                                                                                       like
                                                                                                                                       dark matter\end{tabular}  & 300-2500                                                                & $\times$                                                                  \\\multicolumn{1}{|l|}{}          
 & \begin{tabular}[c]{@{}c@{}}nuclear \\ magnetar\end{tabular} &
                                                                 $\checkmark$                                                                &                                                                  10$^{3-5}$                                              & -3/2                                                                                   & near black hole  & 10-3000                                                                 & $\checkmark$                                                        \\
\multicolumn{1}{|l|}{}
  & SNR pulsar                             &   $\checkmark$                                                               & $10^{1-3}$                                               & -3/2                                                                                   & \begin{tabular}[c]{@{}c@{}}archival SNe \\or SNR \end{tabular}                  & 10$^2$-10$^4$                                                           & $\checkmark$                                                        \\ \cline{1-1}
\multicolumn{1}{|l|}{Galactic ($\lesssim 100$ kpc)}                                                                                
& flaring MS stars       & $\checkmark$     & RM$_{\rm gal}$                                &$>-3/2$                                                                                   & \begin{tabular}[c]{@{}c@{}}main sequence \\ star\end{tabular}               & $\gtrsim$ 300                                                           & $\times$                                                                  \\ \cline{1-1}
\multicolumn{1}{|l|}{Terrestrial ($\lesssim 10^5$ km)}
  & RFI                         &\checkmark  & $\lesssim$ 1                                & ? &                                       diurnal variation                                                                       & ?                                                                       & $\times$                                                                  \\ \hline
\end{tabularx}
\caption{\scriptsize This table summarizes a number of FRB models by classifying them as cosmological, 
extragalactic but local, Galactic, and terrestrial. 
The seven columns are potential observables of FRBs and each
 row gives their consequence for a given model 
 (Blitzars: \protect\cite{2014A&A...562A.137F}), 
 compact object mergers \protect\citep{2012ApJ...760...64M, 2013PASJ...65L..12T},
bursts from magnetars \protect\citep{2014MNRAS.442L...9L}, 
dipole dark matter \protect\citep{2017ApJ...844..162T},
cosmic strings \protect\citep{2017arXiv170702397B},
superradiance \protect\citep{2018MNRAS.475..514H},
edge-on disk galaxies \protect\citep{2015RAA....15.1629X}, 
 exploding primordial blackholes \protect\citep{2014PhRvD..90l7503B},  
circumnuclear magnetars \protect\citep{2015ApJ...807..179P}, 
 supernova remnant pulsars, stellar flares \protect\citep{2014MNRAS.439L..46L}, and terrestrial RFI 
 \protect\citep{2015arXiv150305245H}.).
The parameters are generic expectations of models, and lying outside
this range does not necessarily rule them out.
Even though all models have to explain the observed 180-2,600 pc cm$^{-3}$, some models predict a wider 
range of DM. For instance, in the circumnuclear magnetar or edge-on disk disk scenarios there 
ought to be bursts at relatively low DM that simply have not been identified as FRBs.
}
\label{TAB-1}
\end{table*}
\end{landscape}

\section{Conclusions}

The FRB phenomenon has stimulated broad ideas to generate the short-time energetic coherent radio emission observed.  All FRB's share the
coherence phenomenon.  The repeating FRB121102 presents many
challenges, hinting to activity near a black hole inferred from the
high RM -- no other locale is known to have such a high RM. Magnetic
dipoles, and neutron stars are viable candidates.  The dipole dark
matter model is the only one which makes predictions on properties,
including polarization angle, spectral index, etc.  Resorting to known
unknowns is required in pulsar interpretations, since the pulsar radio
emission mechanism itself is still not understood.

More than one category of sources may be responsible: in the history of gamma-ray bursts (see the Comment by S. R. Kulkarni in this issue), some repeated, while most do not.  Similarly, FRB show a wide
variety of properties, and thus might not all be the same. 

Some models may permit ready counterpart tests if localized through
very long baseline interferometry within their host galaxy: nuclear, supernovae (remnants). Localization of the burst to its nearest galaxy will discriminate between broad distance classes, though it is unlikely to pinpoint any specific model.

Independent of their physical nature, the coherence of FRBs is assumed to
provide us with a rich new probe of our universe: as we spend billions
building synchrotron light sources on earth, FRB are much more
coherent and powerful, enabling true interferometry through their
lensed multipath propagation. The burst path lengths of $10^{25}$m are
probed to $10^{-3}$m, corresponding to a dimensionless strain of
$10^{-28}$, potentially the most precise measurement available.  It
could probe space-time, from lensing\citep{2016PhRvL.117i1301M} to
gravitational waves\citep{2018MNRAS.479..406R} to deviations from
gravity\citep{2017PhRvD..95h4049Y}.

\section{Acknowledgements}

I thank Liam Connor for helpful feedback and NSERC for support.

\bibliographystyle{nature}

\end{document}